\documentclass[preprint,aps,pra,showemail,showpacs]{revtex4-1}

\usepackage{amsthm}
\usepackage{amsmath}
\usepackage{latexsym}
\usepackage{amsfonts}
\usepackage{amssymb}
\usepackage{color}
\usepackage{bbm,dsfont}
\usepackage{graphicx}
\usepackage{subcaption}
\usepackage[pdfpagelabels=true]{hyperref}

\usepackage{tikz}
\usepackage{placeins}
\usepackage[utf8]{inputenc}
\usepackage{braket}

\newtheorem{proposition?}{Proposition?}

\theoremstyle{definition}

\newcommand{\real}{\mathbb R} 
\newcommand{\complex}{\mathbb C} 
\newcommand{\ii}{\mathrm{i}} 

\newcommand{\hi}{\mathcal{H}} 
\newcommand{\tr}[2]{\text{tr}_{#2}\left\{#1\right\}}

\newcommand{\id}{\mathbbm{1}} 

\renewcommand{\L}{\mathcal{L}}

\newcommand{\diff}{{\rm d}}

\begin{document}
\title{Phase space theory for open quantum systems with local and collective dissipative processes}

\author{Konrad Merkel}
\affiliation{Institut f{\"u}r Theoretische Physik, Technische Universit{\"a}t Dresden, 
D-01062, Dresden, Germany}

\author{Valentin Link}
\affiliation{Institut f{\"u}r Theoretische Physik, Technische Universit{\"a}t Dresden, 
D-01062, Dresden, Germany}

\author{Kimmo Luoma}
\email{kimmo.luoma@tu-dresden.de}
\affiliation{Institut f{\"u}r Theoretische Physik, Technische Universit{\"a}t Dresden, 
D-01062, Dresden, Germany}

\author{Walter T. Strunz}
\email{walter.strunz@tu-dresden.de}
\affiliation{Institut f{\"u}r Theoretische Physik, Technische Universit{\"a}t Dresden, 
D-01062, Dresden, Germany}

\date{\today}

\begin{abstract}
In this article we investigate driven dissipative quantum dynamics of an ensemble of two-level systems given by a Markovian master equation with collective and non-collective dissipators. Exploiting the permutation symmetry in our model, we employ a phase space approach for the solution of this equation in terms of a diagonal representation with respect to certain generalized spin coherent states. Remarkably, this allows to interpolate between mean-field theory and finite system size in a formalism independent of Hilbert-space dimension. Moreover, in certain parameter regimes, the evolution equation for the corresponding quasiprobability distribution resembles a Fokker-Planck equation, which can be efficiently solved by stochastic calculus. Then, the dynamics can be seen as classical in the sense that no entanglement between the two-level systems is generated.
Our results expose, utilize and promote techniques pioneered in the context of laser theory, which we now apply to problems of current theoretical and experimental interest.
\end{abstract}

\maketitle

\section{Introduction}
Quantum optical models, such as the Dicke model, where  an ensemble of 
identical two-level atoms is coupled to 
a cavity mode, have been among the first systems  where emergent
collective behavior such as super-radiance has been investigated~\cite{Dicke1954}.  
The cooperative effects can be described in terms of collective atomic 
operators in such a way that the ensemble of $N$ two-level
atoms is treated as a large spin of length $j=\frac{N}{2}$. Interestingly, such a cooperative 
state can still possess  non-trivial quantum correlations~\cite{Dicke1954}. 
The proposal on how to implement  
the Dicke model in an optical cavity QED system~\cite{Dimer2007}
and the experimental realization using a super-fluid gas trapped inside an optical cavity~\cite{Baumann2010} 
are important milestones, which have inspired more studies on the 
role of collective effects in quantum phase transitions, 
such as~\cite{PhysRevLett.123.207402,PhysRevLett.123.243602}.

Purely collective dynamics cannot always be achieved due to experimental
conditions, for example when the ensemble of two level systems 
is inhomogeneous~\cite{PhysRevA.88.043806} or when the individual 
two level systems couple to different reservoirs~\cite{Galve2017}. In these
cases, the dimensionality of the relevant Hilbert space in general grows
exponentially with the system size. An interesting alternative regime exists when in addition to collective 
processes only local processes occur which preserve permutation invariance. This holds if all non-collective processes are identical for each constituent in the ensemble. In this 
regime, permutation invariance can be utilized to reduce the effective
dimension of the problem~\cite{PhysRevA.98.063815}. Formally, this can be
understood as follows. In an ensemble of $N$ two-level systems, collective processes confine the dynamics 
on a subspace $\hi_j\subset\complex^{2N}$ spanned by the Dicke states
$\ket{j,m}$, where $-j \leq m \leq j$, whereas the permutation invariant 
local processes couple the different subspaces $\hi_{j'}$ spanned
by Dicke ladders labeled by differing $j'$. The reduction
in the effective dimension emerges due to high degeneracy of the Dicke states~\cite{mandel1995optical}.

When local and collective processes coexist, a host of interesting phenomena
can be studied. For example, incoherent pumping on a super-radiant ensemble
of two level systems can lead to robust steady state super-radiance~\cite{PhysRevA.81.033847}. 
Competing collective phenomena, such as driving and dissipation can lead to non-equilibrium 
phase transitions~\cite{Iemini2018,PhysRevA.99.062120} and recently the robustness 
of such transitions against local dephasing has been investigated~\cite{Tucker_2018}.

In this article we will investigate a paradigmatic driven, dissipative open system consisting of $N$ two level atoms which is affected by both  
collective and permutation invariant local dissipative processes. We describe the 
dynamics of such a system with the following GKSL type master equation~\cite{Gorini1976,Lindblad1976}
\begin{align}\label{eq:model}
  \partial_t\rho=\L[\rho] = -i[H,\rho]+\L_l[\rho]+\L_c[\rho]\,,
\end{align}
where collective driving is given by 
\begin{align}\label{eq:driving}
  H = \sum_{\lambda=1}^N\frac{1}{2}\vec{B}\cdot\vec\sigma_\lambda\equiv \vec{B}\cdot\vec{J}\,,
\end{align}
with $\vec{B}\in\real^3$, $\sigma_{\lambda}^i=\sigma_i$
for $i=z,\pm$  are Pauli spin matrices $\sigma_z$, 
$\sigma_\pm = \sigma_x \pm i \sigma_y$  acting  on system $\lambda$, and 
$\vec{J}$ is the collective spin operator with components $J_z$ and $J_\pm = J_x \pm i J_y$.
The collective dissipation is given by 
\begin{align}\label{eq:coll_diss}
  \L_c[\rho] =\sum_{k=z,\pm}\L_c^k[\rho]= \sum_{k=z,\pm}\frac{2\kappa_k}{N}\left(J_k\rho J_k^\dagger 
  -\frac{1}{2}\left\{J_k^\dagger J_k,\rho\right\}\right)\,,
\end{align}
whereas the permutation invariant local dissipation is given by
\begin{align}\label{eq:loc_diss}
  \L_s[\rho] = \sum_{k=z,\pm}\sum_{\lambda=1}^N\frac{\gamma_k}{2}
  \left(\sigma_\lambda^k\rho \sigma_\lambda^{k\dagger} 
  -\frac{1}{2}\left\{\sigma_\lambda^{k\dagger} \sigma_\lambda^k,\rho\right\}\right)\,.
\end{align}
We assume that all of the local and collective rates, $\gamma_k$ and $\kappa_k$
respectively, are non-negative, so that the dynamics generated by the master equation is completely positive. Note that the non-collective term does not commute with the total angular momentum $\vec{J}^2$. It therefore couples different eigenspaces of $\vec{J}^2$ in contrast to the collective terms. 
 
Instead of using the basis of Dicke states directly, we will employ a
phase space approach by representing the state of the open system
in terms of generalized
spin coherent states and the associated $P$-function~\cite{Gisin1992,Puri1980}.  
This type of approach to collective 
phenomena has been  already 
discussed in the quantum optics community during the 70s and 80s of the previous century, in
the context of cooperative fluorescence
\cite{Walls1978,Walls1980,Drummond1978,Carmichael1980}. 
The generalized coherent state approach that we take has been 
pioneered in~\cite{Weidlich1967Aug,Weidlich1967Jun,Haken1967Aug} and used also 
more recently in~\cite{Patra2019Mar}. Here we show that in certain parameter regions this method can be used to map the solution of master equation (\ref{eq:model}) to a Fokker-Planck equation for any system size. Then the dynamics is classical in the sense that no quantum correlations between the two-level systems build up. In the method the system size is merely a parameter determining the strength of diffusion in phase space. Therefore, it allows to analytically interpolate between finite-size and mean-field theory. 

The outline of the article is the following. First we introduce a generalization of spin coherent states in section \ref{sec:Generalized_Spin_Coherent_States}. In section \ref{sec:Equation_of_motion_for_the_P-function} we derive the equation of motion for the associated $P$-function. The phase space approach provides immediately a consistent mean field theory, which we explore in section \ref{sec:Mean_field_theory}. In section \ref{sec:semi_class} we focus on the exact semi-classical regime of our model, where the equations of motion can be efficiently solved. At last, we conclude with discussion and outlook in section \ref{sec:discussion}.

\section{Generalized Spin Coherent States} \label{sec:Generalized_Spin_Coherent_States}

The main technical tool of this paper is to expand the state of all
identical two level systems in terms of a generalized
$P$-representation. More precicely, consider the density operator
$\alpha(\vec{r})$ of the simplest product state where all two level
systems are identical
\begin{equation}
\alpha(\vec{r})=\bigotimes\limits_{\lambda=1}^N\frac{1}{2}(\id+\vec{\sigma}\cdot\vec{r})\,. \label{eq:def_alpha}
\end{equation} This state is positive iff the Bloch vector $\vec{r}$
has a length smaller or equal to 1. It is pure iff
$|\vec{r}|^2= 1$. We show later on that in this case the definition~(\ref{eq:def_alpha})
reduces to the standard spin coherent states in the symmetric Dicke
subspace with $j=\frac{N}{2}$. Due to this property, we call these
states the generalized coherent states. However, we stress that these are
not pure states and do not have the same group theoretical 
interpretation  as regular coherent states~\cite{RevModPhys.62.867}. 
We will later express the
state in terms of a diagonal representation using a $P$-function  associated with 
these generalized
coherent states. Firstly, let us discuss how certain operators act on
these states. A direct calculation shows that 
\begin{align}
\sigma_k(\id+\vec{\sigma}\cdot\vec{r})=\big[r_k+(\delta_{ki}-\ii
\varepsilon_{kij}r_j-r_kr_i)\partial_i\big](\id+\vec{\sigma}\cdot\vec{r})\,,\\
(\id+\vec{\sigma}\cdot\vec{r})\sigma_k=\big[r_k+(\delta_{ki}+\ii
\varepsilon_{kij}r_j-r_kr_i)\partial_i\big](\id+\vec{\sigma}\cdot\vec{r})\,,
\end{align} where the derivatives are with respect to $\vec{r}$,
i.e. $\partial_i=\partial/\partial r_i$, and summation over repeated
indices is implied. With these relations we can replace the
action of operators acting on generalized coherent states by
differential operators. One example is the total angular momentum
operator $\vec{J}$
\begin{equation}
\begin{split} J_k\alpha(\vec{r})&=\frac{1}{2}\sum_{\lambda=1}^N
\frac{1}{2}(\id+\vec{\sigma}\cdot\vec{r})\otimes...\otimes
\underbrace{\sigma_k
\frac{1}{2}(\id+\vec{\sigma}\cdot\vec{r})}_{\text{$\lambda$'th
position}} \otimes...\otimes
\frac{1}{2}(\id+\vec{\sigma}\cdot\vec{r})=\text{(use product rule)}\\
&=\frac{N}{2} r_k\alpha(\vec{r})+\frac{1}{2}(\delta_{ki}-\ii
\varepsilon_{kij}r_j-r_kr_i)\partial_i\alpha(\vec{r})\,.
 \end{split}
\end{equation} This gives us the algebra of spin coherent states. It
is analogous to well known relations such as $a\ket{z}=z\ket{z}$ in
the case of bosons. With these rules we can express the action of the
Hamiltonian part, as well as the collective dissipative part of the
Lindbladian (\ref{eq:model}), onto a coherent state as a differential operator with
respect to the coherent state label $\vec{r}$. Due to our more general
definition of spin coherent states, this is also true for the local
dissipators. Consider for instance identical local dephasing (in the $\sigma_z$ 
basis with rate $\gamma_z/2$) of all
two level systems
\begin{equation}
\begin{split}
\mathcal{L}_s^z[\alpha(\vec{r})]=\frac{\gamma_z}{2}\sum_\lambda
\mathcal{L}_{s,\lambda}^z[\alpha(\vec{r})]=\frac{\gamma_z}{2}\sum_{\lambda=1}^N
\frac{1}{2}(\id+\vec{\sigma}\cdot\vec{r})\otimes...\otimes
\underbrace{\mathcal{L}_{\lambda}^z[\frac{1}{2}(\id+\vec{\sigma}\cdot\vec{r})]}_{\text{$\lambda$'th
position}} \otimes...\otimes \frac{1}{2}(\id+\vec{\sigma}\cdot\vec{r})\,.
 \end{split}
\end{equation} Now the action of any superoperator on a two level
system can always be written as a first order differential operator
with respect to the Bloch vector. This is obvious because the state
itself is linear in $\vec{r}$. In the example of dephasing we find
\begin{equation}
\mathcal{L}_{\lambda}^z[\frac{1}{2}(\id+\vec{\sigma}\cdot\vec{r})]
=\sigma_z\frac{1}{2}(\id+\vec{\sigma}\cdot\vec{r})\sigma_z-\{\sigma_z^2,\frac{1}{4}(\id+\vec{\sigma}\cdot\vec{r})\}=(-
2x\partial_x -2y\partial_y)\frac{1}{2}(\id+\vec{\sigma}\cdot\vec{r})\,.
\end{equation} with canonical notation $r_1=x,\, r_2=y,\, r_3=z$.  Inserting this
in the above expression and using the product rule one finds for the coherent state
\begin{equation}
\begin{split} \mathcal{L}_s^z[\alpha(\vec{r})]=(- {\gamma_z}x\partial_x
-{\gamma_z}y\partial_y)\alpha(\vec{r})\,.
 \end{split}
\end{equation} Of course, the same calculations can be done for local
decay and pump channels. In summary, we can now express the action of
the entire Lindblad superoperator on a generalized spin coherent
state as a differential operator with respect to the state label, that
is the vector $\vec{r}$. \\ 
Let us briefly make the connection to the
standard spin coherent states. These are usually defined in an
eigenspace of $\vec{J}^2$ with eigenvalue $j=\frac{N}{2}$ (symmetric
Dicke states). They are rotations of the 'fully polarized' state
$\alpha(\vec{e}_z)$. We recover these states from our generalized
coherent states by setting $|\vec{r}| = r =1$. For better illustration, let
us parameterize the vector with spherical coordinates
\begin{equation}\label{eq:sph_coor} \vec{r}=\begin{pmatrix} r \sqrt{1-\eta^2}\cos\phi\\ r
\sqrt{1-\eta^2}\sin\phi\\ r \eta
         \end{pmatrix}\,,\qquad \phi\in [0,2\pi)\,,\qquad \eta\in
[-1,1]\,,\qquad r\in[0,1]\,,
\end{equation} where one may identify $\eta=\cos\theta$. The differential representation of the dissipators in these coordinates are provided in appendix \ref{appendix:spherical}. The main insight is that for $r=1$ in the
collective operators, no derivatives with respect to $r$ appear. The
radius is preserved because the dynamics does not lead out of the symmetric $j=N/2$
subspace. This is different in the case of local dissipation. The
local dissipators are not confined to this subspace which is reflected in
the occurence of derivatives with respect to $r$. If the dynamics
remains in the Dicke subspace however, the equations of motions which
we derive in the next section will automatically conserve $r=1$ so
that both cases are covered in the formalism. Now as an ansatz we
express the state with a diagonal representation in terms of the
generalized coherent states
\begin{equation} \rho=\int\limits_{r\leq
1}d^3r\,P(\vec{r})\,\alpha(\vec{r})\,,
 \label{eq:rho_ansatz}
\end{equation} with a generalized  $P$-function $P(\vec{r})$, which is a
quasiprobability distribution on the unit ball in $\real^3$ centered
at the origin. As
$\alpha(\vec{r})$ is a polynomial in $\vec{r}$ of order $N$, it is
obvious that $P$ is not unique. If $\rho$ is a normalized state, the
distribution is normalized as well
\begin{equation} 1=\tr{\rho}{}=\int\limits_{r\leq
1}d^3r\,P(\vec{r})\,\tr{\alpha(\vec{r})}{}=\int\limits_{r\leq 1}d^3r\,P(\vec{r})\,.
\end{equation} If $P(\vec{r})$ is a positive function, the state is also positive
by construction. However, $P(\vec{r})$ does not need to be positive. In fact,
if $P(\vec{r})$ is positive for all $\vec{r}$, then the state is
separable by definition, as it is a classical mixture of separable
states. An entangled state necessarily has a negative $P$
function. Once the $P$ function of the state is known, all observables can
be computed easily
\begin{equation} \tr{\rho O}{}=\int\limits_{r\leq
1}d^3r\,P(\vec{r})\,\tr{\alpha(\vec{r})O}{}\,.
\end{equation} For example, in the case of the angular momentum
operator one finds $\tr{\alpha(\vec{r})\vec{J}}{}=N/2\vec{r}$ so that
\begin{equation} \tr{\rho\vec{J}}{}=\frac{N}{2}\int\limits_{r\leq
1}d^3r\,P(\vec{r})\,\vec{r}\,.
\end{equation}

\section{Equation of motion for the P-function} \label{sec:Equation_of_motion_for_the_P-function}

With the differential form of operators at hand, it is a straightforward 
task to derive an equation of motion for the $P$-function of a state evolving
under the GKSL equation~(\ref{eq:model}). One simply plugs in the
expression (\ref{eq:rho_ansatz}) in the equation of motion
\begin{equation}
 \begin{split}
  \partial_t\rho&=\int\limits_{r\leq1}d^3r\,\partial_tP(\vec{r})\,\alpha(\vec{r})
  =\mathcal{L}[\rho]=\int\limits_{r\leq1}d^3r\,P(\vec{r})\,\mathcal{L}[\alpha(\vec{r})]\,.
 \end{split}
\end{equation} Using the differential representation, we can express the action of
the GKSL-generator as a second order differential operator
\begin{equation}
\mathcal{L}[\alpha(\vec{r})]=\Big(a_i(\vec{r})\partial_i+\frac{1}{2N}D_{ij}(\vec{r})\partial_i\partial_j\Big)\alpha(\vec{r})\,.
\end{equation} 
To find the ``forward in time'' equation of motion for $P$ one
performs partial integration under the integral, neglecting boundary
terms. Comparison of the integrands on the left and right hand sides
gives the following partial differential equation
\begin{equation}
 \partial_tP(\vec{r})=\Big(-\partial_ia_i(\vec{r})+\frac{1}{2N}\partial_i\partial_jD_{ij}(\vec{r})\Big)P(\vec{r})\label{eq:FPE}\,.
\end{equation} General expressions for $\vec{a}(\vec{r})$ and
$D(\vec{r})$ are provided in appendix \ref{appendix:xyz}. This exact equation has the form
of a second order Kramers-Moyal expansion. 
However, we point out that  the matrix $D(\vec{r})$
is in general not positive for all $\vec{r}$. When $D(\vec{r})$ is 
positive semidefinite for all $\vec{r}$, the equation corresponds to a
Fokker-Planck equation. Since the Fokker-Planck
equation always preserves the positivity, it is clear
that the diffusion can be positive only if the dynamics does not
generate entanglement between the two level systems. Indeed, we do
find certain parameter regimes in our model where this holds
true. Then the partial differential equation is well behaved and can
easily be integrated numerically using  stochastic
differential equations. We give a detailed description of this in section \ref{sec:semi_class}. \\

\section{Mean field theory} \label{sec:Mean_field_theory}
\begin{figure}
	\begin{subfigure}[l]{0.3\textwidth}
		\centering
		\includegraphics[width=\textwidth]{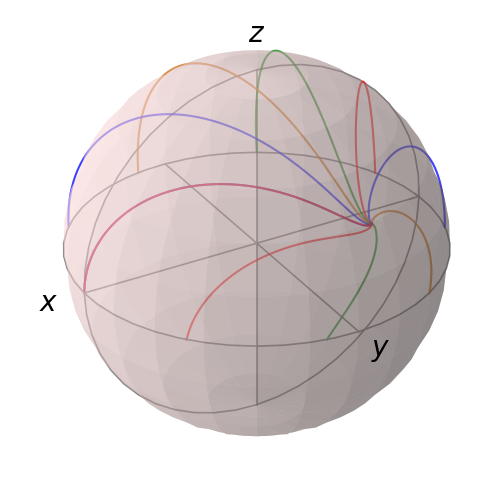}
		\subcaption{purely collective} 
	\end{subfigure}
	\begin{subfigure}[c]{0.3\textwidth}
		\centering
		\includegraphics[width=\textwidth]{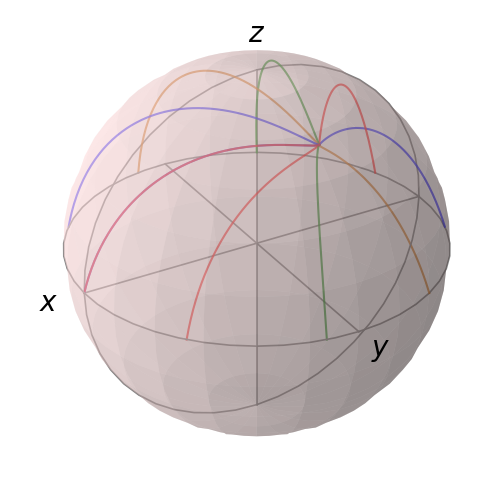}
		\subcaption{local decay $\gamma_+=0.2 B$} 
	\end{subfigure}
	\begin{subfigure}[r]{0.3\textwidth}
		\centering
		\includegraphics[width=\textwidth]{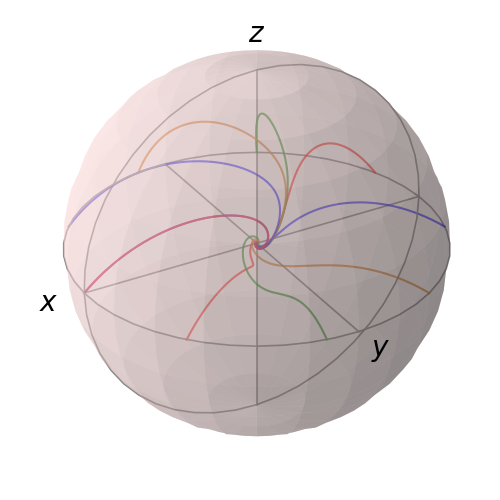}
		\subcaption{local dephasing $\gamma_z=0.4 B$} 
	\end{subfigure}
	\caption{Solutions of the mean-field equations (\ref{eq:mean_field}) depicted on the unit ball. As collective parameters we choose $\vec{B}=B/\sqrt{2}(\vec{e}_z-\vec{e_x})$, $\kappa_+=0.8 B$ and $\kappa_-=0$. (a) Only collective processes are present $\gamma_i=0$ so that the trajectories stay on the surface of the ball. (b) Due to the local decay, the stable fixed point lies inside the ball. (c) If local dephasing is present, all trajectories approach the fully mixed state $r=0$.} \label{fig:mean_field}
\end{figure}

From equation (\ref{eq:FPE}) follows
directly a 'mean field' approximation by neglecting the diffusion
term. In general, this approximation is exact in the thermodynamical
limit $N\rightarrow\infty$ because then the diffusion vanishes exactly
due to the prefactor $\frac{1}{N}$. The remaining equation of motion
is first order and can be solved by the method of characteristics
\begin{equation}
  \partial_tP(\vec{r})=-\partial_ia_i(\vec{r})P(\vec{r})\label{eq:Liouville}\,.
\end{equation} The solution is given by the time evolution of the
initial distribution along the mean-field trajectories
\begin{equation} P_t(\vec{r})=\int\diff^3r_0
P_0(\vec{r}_0)\delta(\vec{r}-\vec{r}(t))\,,\qquad
\vec{r}(0)=\vec{r}_0\,,\qquad \dot{\vec{r}}(t)=\vec{a}(\vec{r}(t))\,.
\end{equation} If the system starts in a coherent state,
i.e. $\rho_0=\alpha(\vec{r}_0)$ or
$P_0(\vec{r})=\delta(\vec{r}-\vec{r}_0)$, then the state remains a
coherent state under mean field evolution and
$\rho_t=\alpha(\vec{r}(t))$ or
$P_t(\vec{r})=\delta(\vec{r}-\vec{r}(t))$. For our master equation the
mean field equations of motion are explicitly
\begin{equation}\label{eq:mean_field}
\begin{split} &\dot{x}= {B_y} z-{B_z} y+x
(\gamma_--\gamma_+-\gamma_z+\kappa_- z-\kappa_+ z)\,,\\
&\dot{y}=-{B_x}
z+{B_z} x+y (\gamma_--\gamma_+-\gamma_z+\kappa_- z-\kappa_+ z)\,, \\
&\dot{z}={B_x} y-{B_y} x-(\kappa_--\kappa_+) \left(x^2+y^2\right)+2
\gamma_- (z-1)-2 \gamma_+ (z+1)\,,
\end{split}
\end{equation} where we have neglected subleading terms of order
$1/N$ consistent with neglection of the diffusion. Note that
collective dephasing does not influence the mean field
dynamics. Except for some fine tuned cases
\cite{PhysRevA.99.062120,Hannukainen2018}, mean field theory
characterizes the steady state phases of the model
\cite{Ferreira2019}. Phase diagrams can be derived by finding the
fixed points of (\ref{eq:mean_field}) and analyzing their
stability. We show a few examples of solutions of the mean-field equation in Fig.~\ref{fig:mean_field}. The images give an intuitive picture of the influence of the different dissipators onto the mean-field solutions.

\section{Exact semi-classical regime}\label{sec:semi_class}
\begin{figure}
	\centering
	\begin{subfigure}[l]{0.49\textwidth}
		\centering
		\includegraphics[width=\textwidth]{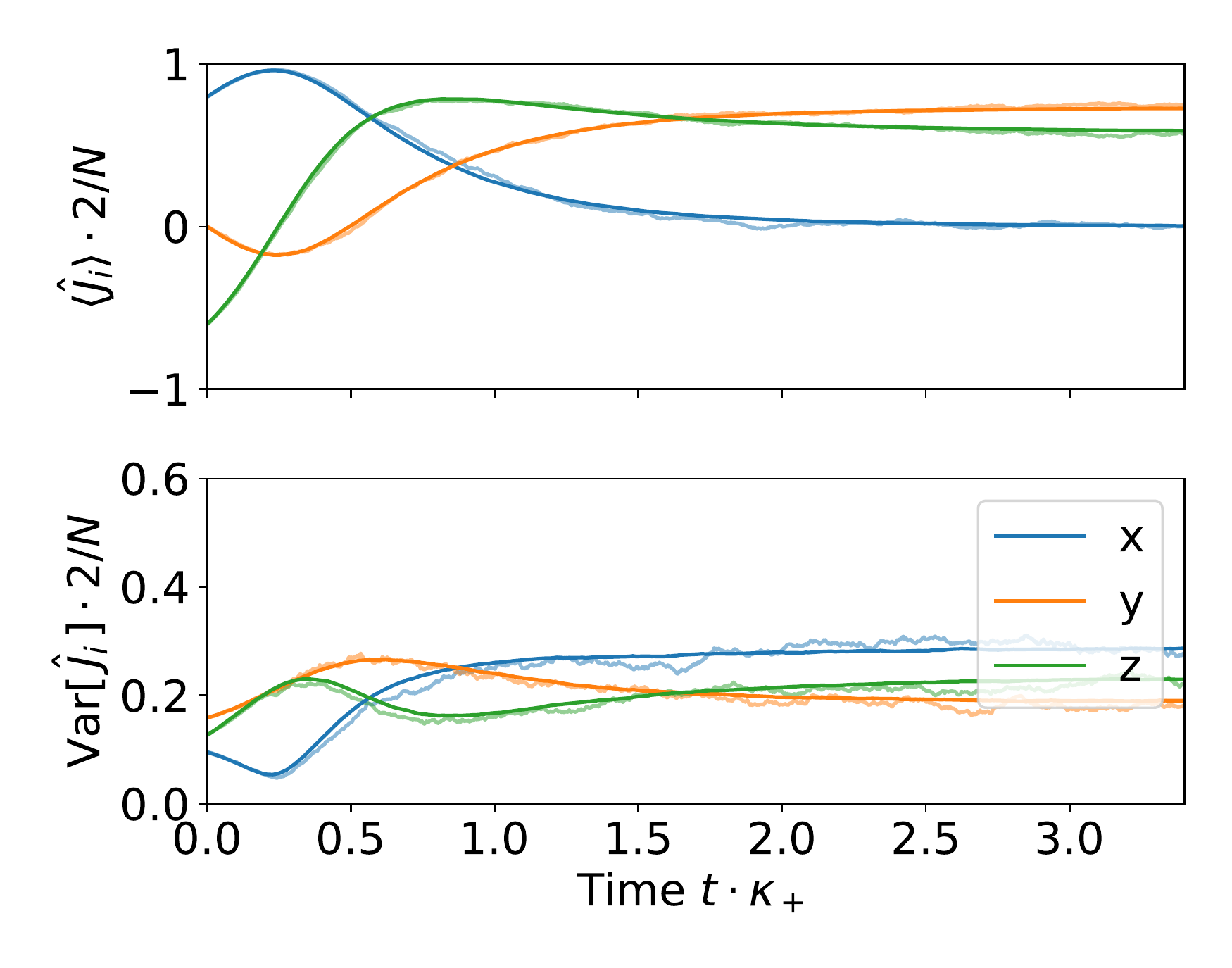}
		\subcaption{$B_x = -0.8\kappa_+$} \label{fig:result_01}
	\end{subfigure}
	\begin{subfigure}[r]{0.49\textwidth}
		\centering
		\includegraphics[width=\textwidth]{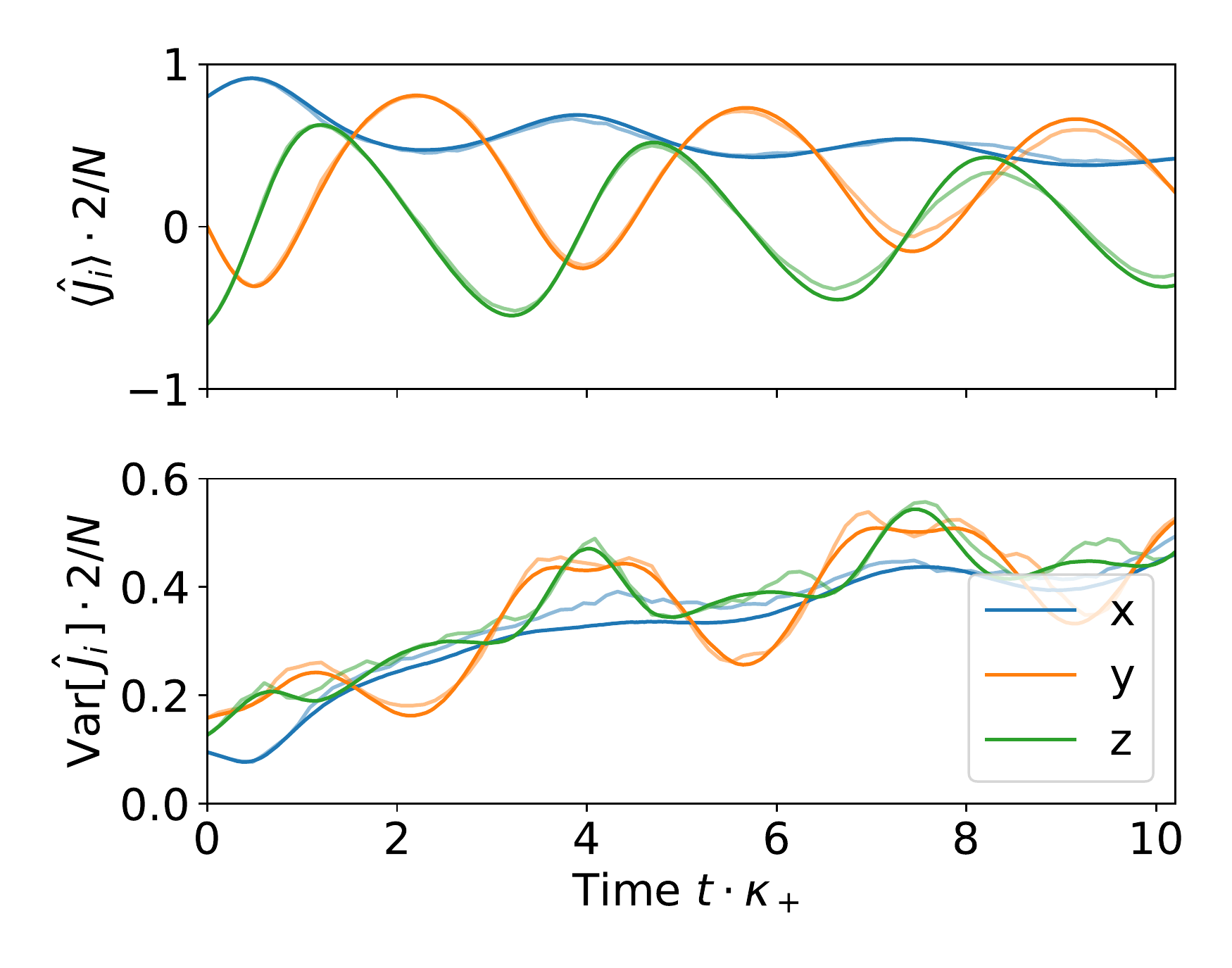}
		\subcaption{$B_x = -2.4\kappa_+$} \label{fig:result_02}
	\end{subfigure}
	\caption{Numerical solution of the master equation (\ref{eq:model})
		with the classical stochastic trajectories obeying
		Eq.~(\ref{eq:Ito}). The light and bold curves are averaged over 100 and 10000 sampled trajectories, respectively. The initial state is a spin coherent state. In
		both figures we choose the parameters $N = 40$, $\kappa_z=\kappa_+$
		and $B_y=B_z=0$. There are no non-collective processes
		$\gamma_{\pm,z}=0$.} \label{fig:trajectories}
\end{figure}
\begin{figure}
	\begin{subfigure}[l]{0.3\textwidth}
		\centering
		\includegraphics[width=\textwidth]{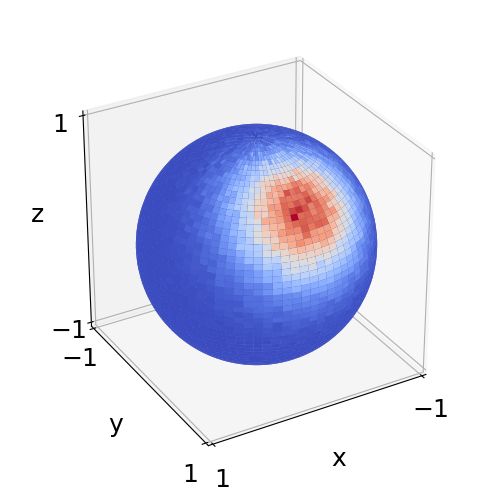}
		\subcaption{$N=20$} 
	\end{subfigure}
	\begin{subfigure}[c]{0.3\textwidth}
		\centering
		\includegraphics[width=\textwidth]{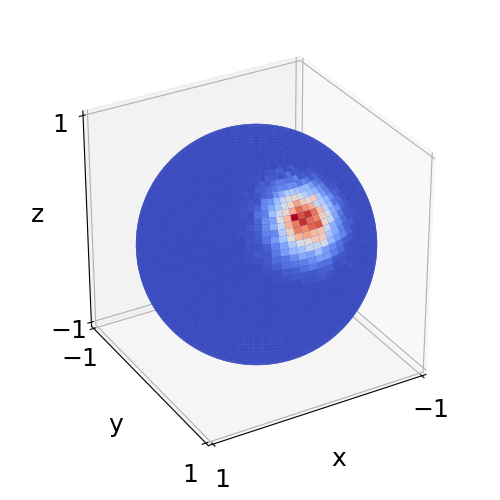}
		\subcaption{$N=100$} 
	\end{subfigure}
	\begin{subfigure}[r]{0.3\textwidth}
		\centering
		\includegraphics[width=\textwidth]{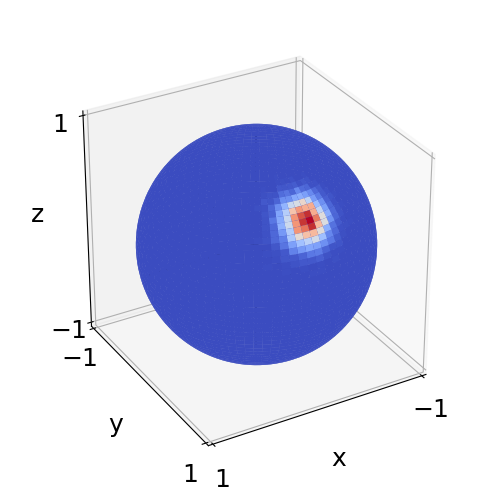}
		\subcaption{$N=200$} 
	\end{subfigure}
	\caption{The pictures show the P-function of the steady state as a
		function on the sphere ($|\vec{r}| = 1$) for three different system sizes $N$. All other
		parameters are the same as in Fig.~\ref{fig:trajectories}~(a). The width of
		the distribution is decreasing when $N$ is
		increasing.} \label{fig:steady_state}
\end{figure}
Going beyond mean-field one must include the diffusion term in the
$P$-function evolution equation. This term is generated by collective dissipation in the master equation (\ref{eq:model}). Using spherical
coordinates, the diffusion matrix reads
\begin{equation}\label{eq:diff_mat}
\left(
\begin{array}{ccc}
 \frac{2\left(1-\eta ^2\right) (\eta  (\kappa_--\kappa_+)+r
   (\kappa_-+\kappa_+))}{r} & 0 & -\left(\eta ^2-1\right)
   \left(r^2-1\right) (\kappa_--\kappa_+) \\
 0 & 2\kappa_z+2\frac{\eta  (\kappa_-(\eta r+1)+\kappa_+ (\eta  r-1))}{\left(1-\eta ^2\right) r} & 0 \\
 - \left(\eta ^2-1\right) \left(r^2-1\right) (\kappa_--\kappa_+) & 0 & 0 \\
\end{array}
\right)\,.
\end{equation}
If the diffusion matrix is not positive semidefinite, the partial differential equation is no
longer parabolic and numerical solution with, for example, finite
element methods, is unstable. There exist parameter regimes where this
matrix has only positive eigenvalues. In this case the dynamics is
described by a Fokker-Planck equation and can be solved with
stochastic methods. An obvious example is the case of dephasing only,
i.e. $\kappa_-=\kappa_+=0$. Dephasing can be realized by a stochastic
unitary evolution~\cite{doi:10.1007/s11080-005-0487-1,PhysRevA.74.022102,mller2019geometric}. 
The corresponding stochastic
Hamiltonian is non-interacting, so that the full dynamics can be
mapped to a classical stochastic process. This is reflected by the
positivity of the diffusion matrix. More generally, whenever the loss and
the pump rates are identical $\kappa_+=\kappa_-$, the diffusion is
positive.\\ Let us now focus on the purely collective case with no
local decay processes. Then the Fokker-Planck equation preserves the
radial direction $r$, so that we can set $r=1$ and recover the
canonical spin coherent state $P$-function. Then, the matrix~(\ref{eq:diff_mat})
has two non-zero eigenvalues given by
\begin{align}
\lambda_1 &=  2(1-\eta^2)((1+\eta)\kappa_-+(1-\eta)\kappa_+)\,,\\
\lambda_2 &= 2\kappa_z+\kappa_-\frac{2\eta}{1-\eta}-\kappa_+\frac{2\eta}{1+\eta}\,.
\end{align}
We see that Eq.~(\ref{eq:FPE}) is a proper Fokker-Planck equation if 
both $\lambda_1,\lambda_2$ are positive. $\lambda_1$ is always positive since $-1\leq\eta\leq 1$
and $\lambda_2$ is positive for all $\eta$ and $\phi$ if
\begin{align}
 \kappa_++\kappa_-\leq 2\kappa_z\,.
 \label{eq:condition_for_pos_diffusion}
\end{align}
Satisfying this condition,  the open quantum system evolution described by Eq.~(\ref{eq:model}) 
can be mapped to a classical diffusion process on the
sphere. Nevertheless the state lies in a finite-dimensional
Hilbert-space and contains quantum fluctuations due to the non-zero
overlap of coherent states. Relation~(\ref{eq:condition_for_pos_diffusion}) can be understood in the sense that one simply has to add enough dephasing to compensate the positivity violation of the diffusion matrix due to collective losses. We point out that adding dephasing does however not influence the mean-field theory and thus the different phases of the model.   \\
The standard way of solving the Fokker-Planck equation is to consider the 
set of stochastic differential equations
\begin{align} \label{eq:Ito}
\diff\eta &= a_\eta \diff t+ \sqrt{\frac{\lambda_1}{N}} \diff\xi_\eta\,, \\
\diff\phi &= a_\phi \diff t+ \sqrt{\frac{\lambda_2}{N}} \diff\xi_\phi\,,
\end{align}
where $\diff \xi_i$ is a real Ito increment with properties 
$\diff\xi_i\diff\xi_j=\delta_{ij}\diff t$ and the drift terms are given in appendix \ref{appendix:spherical}.
In the absence of the noise terms we obtain again the mean-field theory, as in the last section. 
The stochastic equations provide an efficient way of calculating the $P$-function of the
system for \emph{any} system size $N$ when condition
(\ref{eq:condition_for_pos_diffusion}) is satisfied. Since in Eq.~(\ref{eq:Ito}) $N$ is merely a parameter determining the strength of
the diffusion, the numerical effort for computing the state with this
stochastic method is independent of the dimension of the underlying
Hilbert-space. 
The method becomes more efficient than direct numerical integration of the master
equation for moderate system sizes, i.e. when the mean field approximation is not yet applicable.
We find that expectation values converge quickly with
respect to the number of sampled trajectories, as we can see from the following example. \\
We consider the cooperative resonance fluorescence model \cite{Walls1978,Walls1980,Drummond1978,Carmichael1980}
with additional collective dephasing. The model is described by the master
equation (\ref{eq:model}) with $\kappa_-=B_y=B_z=0$,
$\kappa_+=\kappa_z$ and without non-collective terms
$\gamma_+=\gamma_-=\gamma_z=0$. It features a mirror symmetry
$J_x\rightarrow -J_x$ which is spontaneously broken when the driving
exceeds the collective damping $|B_x|>\kappa_+$. If the damping is
large, as in Fig.~\ref{fig:trajectories}~(a), a localized steady state
is reached as predicted by mean-field theory. In the symmetry broken
regime the dynamics is characterized by damped oscillations towards a
steady state with large variance, as seen in Fig.~\ref{fig:trajectories}~(b). This is an example where the mean field theory
does not predict correctly the steady state and the finite system size
remains relevant even for large $N$, see Ref.~\cite{PhysRevA.99.062120}
for a more detailed discussion. \\ In Fig.~(\ref{fig:steady_state}) we show
an example of a steady state P-function of the model as a density on
the sphere for different system sizes, in the case of large
damping ($|B_x|>\kappa_+$). As the system size $N$ increases, the distribution gets more
localized due to weaker diffusion. For large system sizes the steady
state distribution is sharply peaked at the stable fixed point of the
corresponding mean-field theory. \\

\section{Discussion} \label{sec:discussion}
In this article we have further developed a particular phase space picture 
for driven dissipative spin systems first introduced in the context of laser 
theory \cite{Haken1967Aug}. Our approach incorporates
both collective and permutation invariant local decay processes into a unified 
framework, which allows for a very simple derivation of evolution equations for the corresponding $P$-function.\\
As is well known, the 
coefficient matrix for the second order terms in the equation of motion for
phase space quasiprobability distributions is typically not positive semidefinite.
We have examined conditions under which 
initially positive phase space densities remain positive for all times. Then the dynamics can be considered classical in the sense that no entanglement is generated between the two-level systems. 
In particular, in the case of purely collective dynamics, presence of sufficiently large dephasing ensures the positivity. Another interesting case is the 'infinite temperature' limit $\kappa_+=\kappa_-$, which results in positive diffusion even in presence of non-collective processes. In case the diffusion matrix is not positive, solving the problem using phase space techniques is challenging as this leads to regions where $P$ is negative and the evolution equation is no longer parabolic.\\
The phase space approach provides 
immediately a consistent mean field picture just by neglecting the 
subleading second order terms in the evolution equation. By analyzing the stability of the fixed points of the 
mean field equations, phase diagrams for non-equilibrium phases can be 
obtained. However, scenarios are known where finite-size corrections become relevant even for large system sizes~\cite{PhysRevA.99.062120}. In the case of positive diffusion, such finite-size corrections can be incorporated exactly by adding noise to the mean-field equation. This remarkable feature allows to efficiently solve the problem with stochastic methods, where the numerical effort is independent of the dimension of the underlying Hilbert space.\\
The formalism presented in this paper can be straightforwardly generalized for $\text{SU}(n)$-type models \cite{Barry2008Nov,Gras2013}, by using generalized Bloch-representations, for example for three-level systems, and defining the coherent states accordingly. \\
We believe that the presentation of the phase space methods in this article is easily accessible and can be applied effortlessly to problems of current experimental and theoretical interest.

\acknowledgements
V.L. is grateful for illuminating discussions with Nathan Shammah and Fabrizio Minganti. V.L. acknowledges support from the international Max-Planck research school (IMPRS) of MPIPKS Dresden. 

\appendix

\section{Drift and diffusion in xyz parameterization}\label{appendix:xyz}
In this appendix we provide the full expressions for the diffusion matrix and the drift term. These can be derived by expressing the action of the generator of master equation (\ref{eq:model}) onto a spin coherent state
\begin{equation}\label{eq:L_on_a}
 \L\alpha(\vec{r})=\Big(\partial_ia_i(\vec{r})+\frac{1}{2N}\partial_i\partial_jD_{ij}(\vec{r})\Big)\alpha(\vec{r})\,.
\end{equation}
Expressing the state via (\ref{eq:rho_ansatz}), this results in an evolution equation for the $P$-function of model (\ref{eq:model})
\begin{equation}
 \partial_tP(\vec{r})=\Big(-\partial_ia_i(\vec{r})+\frac{1}{2N}\partial_i\partial_jD_{ij}(\vec{r})\Big)P(\vec{r})\,.
\end{equation}
The drift term $\vec{a}$ reads
\begin{equation}
\begin{split}
 &a_x={B_y} z-{B_z} y+\frac{x (-\kappa_+-\kappa_z+(N-1) z (\kappa_--\kappa_+))-\kappa_- x}{N}+x (\gamma_--\gamma_+-\gamma_z)
  \,, \\
 &a_y=-{B_x}  z+{B_z}  x+\frac{+y (-\kappa_+-\kappa_z+(N-1) z (\kappa_--\kappa_+))-\kappa_- y}{N}+y (\gamma_--\gamma_+-\gamma_z)  \,,\\
 &a_z={B_x}  y-{B_y}  x+\frac{-2 \kappa_-+2 \kappa_++\kappa_- \left(-(N-1) \left(x^2+y^2\right)-2
   z\right)+\kappa_+ \left((N-1) \left(x^2+y^2\right)-2
   z\right)}{N}\\&\qquad+2 \gamma_- (z-1)-2 \gamma_+ (z+1)  \,,\\
\end{split}
\end{equation}
and the symmetric diffusion matrix $D=D^T$ is given as
\begin{equation}
\begin{split}
 &D_{xx}=2 \left(\kappa_- z \left(-x^2+z+1\right)+\kappa_+ z \left(x^2+z-1\right)+\kappa_z y^2\right) \,, \\
 &D_{xy}=-2 x y (\kappa_z+z (\kappa_--\kappa_+)) \,, \\
 &D_{xz}=x \left(\kappa_- \left(x^2+y^2-(z+1)^2\right)-\kappa_+
   \left(x^2+y^2-(z-1)^2\right)\right) \,, \\
 &D_{yy}=2 \left(\kappa_z x^2+\kappa_- z \left(-y^2+z+1\right)+\kappa_+ z
   \left(y^2+z-1\right)\right) \,, \\
 &D_{yz}=y \left(\kappa_- \left(x^2+y^2-(z+1)^2\right)-\kappa_+
   \left(x^2+y^2-(z-1)^2\right)\right)  \,.\\
 &D_{zz}=2 \left(x^2+y^2\right) (\kappa_-+\kappa_++\kappa_- z-\kappa_+ z)\,.
\end{split}
\end{equation}

\section{Drift and diffusion in spherical parameterization}\label{appendix:spherical}

Parameterizing the spin coherent state with spherical coordinates as in (\ref{eq:sph_coor}), we can transform the derivatives in  (\ref{eq:L_on_a}) accordingly. Expressing the state as
\begin{equation}
 \rho=\int\diff\eta\int\diff\phi\int\diff r \,\alpha(\vec{r})P(\eta,\phi,r)\,,
\end{equation}
the $P$-function obeys an evolution equation of the same form as (\ref{eq:FPE}). In particular
\begin{equation}
 \partial_tP=\Big(-\sum_{\beta={\eta,\phi,r}}\partial_\beta a_\beta+\sum_{\beta,\beta'={\eta,\phi,r}}\frac{1}{2N}\partial_{\beta}\partial_{\beta'}D_{\beta, \beta'}\Big)P \,,
\end{equation}
with drift 
\begin{equation}
\begin{split}
 &a_\eta= \sqrt{1-\eta ^2} ({B_y} \cos (\phi )-{B_x} \sin (\phi ))+2 \gamma_- (\eta  r-1)-2
   \gamma_+ (\eta  r+1)-\frac{1}{2}\left(\eta ^2-1\right) r (\kappa_--\kappa_+)\\ 
   & \qquad+\frac{2
   (\kappa_--\kappa_+)+2 \eta  (\eta  (\kappa_--\kappa_+)+2
   r (\kappa_-+\kappa_+))}{2 N r}\,, \\
 &a_\phi=\frac{-\sqrt{1-\eta ^2} \eta  ({B_x} \cos (\phi )+{B_y} \sin (\phi ))-{B_z}
   \eta ^2+{B_z}}{\eta ^2-1} \,,\\
 &a_r=-2 \eta  (\gamma_-+\gamma_+)+\left(\eta ^2+1\right) r (\gamma_--\gamma_+)+\gamma_z \left(\eta ^2-1\right) r \,,\\
\end{split}
\end{equation}
and diffusion (c.f. Eq.~(\ref{eq:diff_mat}))
\begin{equation}
\begin{split}
 &D_{\eta,\eta}=-\frac{2 \left(\eta ^2-1\right) (\eta  (\kappa_--\kappa_+)+r
   (\kappa_-+\kappa_+))}{r} \,,\\
   &D_{\phi,\phi}=-\frac{2
   \left(\eta  (\kappa_--\kappa_+)+\eta ^2 r (\kappa_-+\kappa_+-\kappa_z)+\kappa_z r\right)}{\left(\eta
   ^2-1\right) r} \,,\\
   &D_{\eta,r}=\left(\eta ^2-1\right)
   \left(-\left(r^2-1\right)\right) (\kappa_--\kappa_+) \,,\\
   &D_{\phi,\eta}=D_{\phi,r}=D_{r,r}=0\,.
\end{split}
\end{equation}

\bibliography{bib}

\end{document}